\documentclass[a4paper,11pt]{article}

\usepackage[preprint,nonatbib]{nips_gibs}

\usepackage[english]{babel}
\usepackage[T1]{fontenc}
\usepackage[utf8]{inputenc}
\usepackage{graphicx}                       % Include figure files
\usepackage{fullpage}
\usepackage[dvipsnames]{xcolor}

\usepackage{listings}

\lstdefinestyle{BashStyle}{
  language=bash,
  basicstyle= \linespread{1.1} \small \sffamily,
  numbersep=3pt,
  keywordstyle=\sffamily,
  commentstyle=\color{gray} \itshape,
  frame=tb,
  columns=fullflexible,
  backgroundcolor=\color{yellow!20},
  linewidth= 0.98\linewidth,
  xleftmargin = 0.02\linewidth
}

\lstdefinestyle{SmallBashStyle}{
  language=bash,
  basicstyle= \linespread{1.1} \footnotesize \sffamily,
  numbersep=3pt,
  keywordstyle=\sffamily,
  commentstyle=\color{gray} \itshape,
  frame=tb,
  columns=fullflexible,
  backgroundcolor=\color{yellow!20},
  linewidth= 0.98\linewidth,
  xleftmargin = 0.02\linewidth
}

\lstdefinestyle{CStyle}{
    language=C++,
    basicstyle=\linespread{1.1} \small \sffamily,
    keywordstyle=\color{RoyalBlue}\sffamily,
    stringstyle=\color{BrickRed}\sffamily,
    commentstyle=\color{ForestGreen}\sffamily,
    morecomment=[l][\color{Plum}]{\#},
    numbersep=3pt,
    frame=tb,
    columns=fullflexible,
    backgroundcolor=\color{yellow!20},
    linewidth= 0.98\linewidth,
    xleftmargin = 0.02\linewidth
}

\lstdefinestyle{SmallCStyle}{
    language=C++,
    basicstyle=\linespread{1.1} \footnotesize \sffamily,
    keywordstyle=\color{RoyalBlue}\sffamily,
    stringstyle=\color{BrickRed}\sffamily,
    commentstyle=\color{ForestGreen}\sffamily,
    morecomment=[l][\color{Plum}]{\#},
    numbersep=3pt,
    frame=tb,
    columns=fullflexible,
    backgroundcolor=\color{yellow!20},
    linewidth= 0.98\linewidth,
    xleftmargin = 0.02\linewidth
}

\lstdefinestyle{SQLStyle}{
  language=SQL,
  basicstyle=\linespread{1.1} \small \sffamily,
    numbersep=3pt,
    frame=tb,
    columns=fullflexible,
    keywordstyle=\color{RoyalBlue} \sffamily,
  backgroundcolor=\color{yellow!20},
  linewidth= 0.98\linewidth,
  xleftmargin = 0.02\linewidth,
  showspaces=false,
  commentstyle=\color{gray}
}

\lstdefinestyle{NoStyle}{
  language=SQL,
  basicstyle=\linespread{1.1} \small \sffamily,
    numbersep=3pt,
    frame=tb,
    columns=fullflexible,
    keywordstyle=\sffamily,
  backgroundcolor=\color{yellow!20},
  linewidth= 0.98\linewidth,
  xleftmargin = 0.02\linewidth,
  showspaces=false,
  commentstyle=\sffamily
}

\lstdefinestyle{SmallNoStyle}{
  language=SQL,
  basicstyle=\linespread{1.1} \footnotesize \sffamily,
    numbersep=3pt,
    frame=tb,
    columns=fullflexible,
    keywordstyle=\sffamily,
  backgroundcolor=\color{yellow!20},
  linewidth= 0.98\linewidth,
  xleftmargin = 0.02\linewidth,
  showspaces=false,
  commentstyle=\sffamily
}

\newcommand{\cod}[1]{\lstinline[style=CStyle]{#1}}
\newcommand{\bcod}
{
\medskip
\begin{lstlisting}[style=CStyle]
}

\newcommand{\bsmallcod}
{
\medskip
\begin{lstlisting}[style=SmallCStyle]
}

\newcommand{\bno}
{
\medskip
\begin{lstlisting}[style=NoStyle]
}

\newcommand{\bsmallno}
{
\medskip
\begin{lstlisting}[style=SmallNoStyle]
}

\newcommand{\bash}[1]{\lstinline[style=BashStyle]{#1}}
\newcommand{\bbash}
{
\medskip
\begin{lstlisting}[style=BashStyle]
}

\newcommand{\sql}[1]{\lstinline[style=SQLStyle]{#1}}
\newcommand{\bsql}
{
\medskip
\begin{lstlisting}[style=SQLStyle]
}

\newcommand{\ecod}
{
\end{lstlisting}
}

\usepackage{enumitem}
\setitemize{noitemsep,topsep=2pt,parsep=2pt,partopsep=2pt}
\setenumerate{noitemsep,topsep=2pt,parsep=2pt,partopsep=2pt}
\setdescription{noitemsep,topsep=2pt,parsep=4pt,partopsep=2pt}

\graphicspath{{./figures/}}

\begin{document}

\author{
Marcelo Mottalli \\
Universidad de Buenos Aires
\And Alejo Sanchez \\
ologan.com
\And Gustavo Ajzenman \\
Grandata, San Francisco
\And Carlos Sarraute \\
Grandata Labs, Buenos Aires
}

\title{Snel: SQL Native Execution for LLVM}

\date{}
\maketitle{}

\begin{abstract}
Snel is a relational database engine featuring Just-In-Time (JIT) compilation of queries and columnar data representation. Snel is designed for fast on-line analytics by leveraging the LLVM compiler infrastructure. It also has custom special methods like resolving histograms as extensions to the SQL language.
``Snel” means “SQL Native Execution for LLVM”.

Unlike traditional database engines, it does not provide a client-server interface. Instead, it exposes its interface as an extension to SQLite, for a simple interactive usage from command line and for embedding in applications.
Since Snel tables are read-only, it does not provide features like transactions or updates. This allows queries to be very fast since they don’t have the overhead of table locking or ensuring consistency.

At its core, Snel is simply a dynamic library that can be used by client applications. It has an SQLite extension for seamless integration with a traditional SQL environment and simple interactive usage from command line.
\end{abstract}

\newpage
\setlength{\parskip}{0pt}

\tableofcontents

\newpage

\setlength{\parskip}{6pt}

% Section 1
%!TEX root = snel.tex

%-----------------------------------------------------
\section{Introduction}
%-----------------------------------------------------

%-----------------------------------------------------
\subsection{A Bit of History}
%-----------------------------------------------------

Snel is a relational database engine featuring Just-In-Time (JIT) compilation of queries and columnar data representation. It is designed for fast on-line analytics by leveraging the LLVM compiler infrastructure~\cite{lattner2004llvm}. It also has custom special methods like resolving histograms as extensions to the SQL language.

What is the motivation for building Snel? 
What motivates the design decisions?
Here is a brief outline of the history of Snel.

\medskip

The R\&D team of the company was working on a product that leverages anonymized datasets coming from mobile phone companies and banks to analyze multiple aspects of the human dynamics of users, such as their consumption patterns, and how these patterns correlate with their economic status~\cite{Leo2016socioeconomic,Leo2017higher,Leo2016correlations}.
The team generated inferences of users' income~\cite{Fixman2016bayesian} and analyzed how income is tied to the location in the communication graph topology~\cite{Luo2017inferring}.
Call networks and calling behavior were also used to infer credit risk~\cite{oskarsdottir2018value} and even lifestyles of users~\cite{DiClemente2017sequence}.

The information concerning the mobility of users was used to predict mobility patterns~\cite{Ponieman2013human} and large social events~\cite{Ponieman2015mobility}. This information can complement traditional urban planning tools~\cite{Anapolsky2014exploracion,Sarraute2015social,Sarraute2015city}.
Universal laws concerning the regularity of users' mobility were observed in \cite{Mucelli2016regularity}.
Mobility was also shown to be correlated with data traffic usage, which in turn was used to characterize users~\cite{Fattori2017new,Oliveira2015measurement,Mucelli2015analysis}.
The mobility information has also applications in epidemiology, providing a cost-effective tool to analyze the spatial spread of diseases such as the Chagas disease~\cite{deMonasterio2016analyzing,Monasterio2017uncovering}.
Finally the mobility information is the input used to generate predictions of the users' location, using methods such as the ones described in \cite{Chen2017spatio,Chen2017towards,Silveira2016mobhet}.

The social graph constructed from the communications between mobile phone users is rich in information that was leveraged to make inferences such as dynamic communities~\cite{Sarraute2013evolution}, inferences of age, gender and other socio-demographic attributes~\cite{Sarraute2014study,Brea2014harnessing,Sarraute2015inference2,Sarraute2015inference,Fixman2016bayesian}. In particular it provides an effective tool for predicting customer churn~\cite{Oskarsdottir2016comparative,Oskarsdottir2017churn,Oskarsdottir2017social}, along with a variety of other predictions~\cite{Chen2016relevance,Mucelli2017mobile,Jouaber2017impact,Fixman2017comparison,Liao2017prepaid}.

\medskip

At that time, the Data Processing Infrastructure (DPI) team was working on a front-end which displays the output of queries made on millions of records containing the results of the previous analyses, and making those queries on a regular MySQL database took minutes, whereas our objective was to give the result to the user in less than a second.

The DPI team came up with the following proposal: to build a tiny engine that would have all the data in memory (since the data could fit in the RAM of the server) and eliminate the I/O latency problems. To simplify things, they decided to:
\begin{enumerate}
\item Use the SQLite engine that has an API to incorporate external data sources.

\item Program everything in C, since this is the language of the SQLite API.

\end{enumerate}

It should be noted that, at that time, the set of queries made by the original product were very limited, so the idea was to hard-code those queries in our engine to accelerate everything.

So the team started working on that. What took them more time was to understand the API of external (or ``virtual'') tables of SQLite. Basically the only thing it did was bring the rows one by one, but luckily it would not bring the whole row, it only brought the columns that were needed, so it was possible to make the engine columnar.
This way the new engine would be a simple  interface to a list of vectors (arrays) where each vector would be a column. Since the SQLite API calls this ``virtual tables'', the name of the engine would be ``virtual table of vectors''. We decided to call it VTor, to shorten it.

%-----------------------------------------------------
\subsection{Our Own SQLite Engine}
%-----------------------------------------------------

When we finished implementing VTor, there was a BIG problem: It was slow. Very slow.

The problem was that the only thing that the SQLite API did was bring the rows one at a time, and the SQLite queries engine was in charge of resolving the queries. 
This is the classical iterator model for query evaluation which was proposed
back in 1974~\cite{lorie1974xrm}, and was made popular by the Volcano
system~\cite{mckenna1993efficient}. It is still today the most commonly used execution strategy, as it is flexible and quite simple. As long as query
processing was dominated by disk I/O the iterator model
worked fine. 
The problem is that, for example, if we have a table with 3 million records, the SQLite API is going to make 3 million calls to the\cod{xNext} function, which is totally useless. We did the calculations and discovered that doing a scan of 1 million rows took approximately 1 second, just to advance the cursor. Without even taking into account the extra cost of processing the query.

As often happens, despair was the mother of inspiration. We got fully into the SQLite code and tried to understand how the queries resolution mechanism worked. It turns out that SQLite has a kind of ``virtual machine'' JVM style called VDBE (Virtual Database Engine), and each query is represented as a series of opcodes that are then interpreted by SQLite to solve the query.

Then we came up with the idea: we could add new instructions to the VDBE that would allow to yield the query execution control, effectively bypassing the SQLite mechanism. In this way we could really hardcode the queries and execute them all within our own engine, while maintaining some compatibility with SQLite. This way we would not need for example to make our own SQL parser since we could ``hook'' the SQLite parser.

So we added new opcodes to VDBE. We called these opcodes ``custom queries'', since they allowed the external engine to execute the query in a customized way. We also extended the SQLite API to allow the external engine to provide more information to the SQLite query planner. The functions that we added were:

\begin{description}
\item[xCanOptimizeSelect:] SQLite passes the tokenized tree of the SQL query to the module and the module responds with a true or false if that particular query can be executed entirely by the engine. If so, a new VDBE opcode named\cod{OP_VTRunSelect} is generated that calls a function of the API.

\item[xExecuteCustomSelect:] It re-passes the tokenized tree to the module but this time it tells it to execute that query. The module returns a cursor that will return a new row each time\cod{xNext} is called.
The difference between this and the ``traditional'' mechanism of the SQLite API is that the module can do anything with the query that is passed to it. What VTor does is to analyze the query and in each call to\cod{xNext} it makes the aggregation of all the data using OpenMP and other optimizations.

\item[xExplainCustomSelect:] It simply asks the module to describe what it will do with the query. It returns a string with a description of the ``query plan'' that the module will execute internally. VTor does not return anything useful, but Snel does, as we will see later.

\end{description}

An additional change we did to make everything run faster was to add indexes. The indices are very simple: they are simply a vector of tuples (value, rowid) ordered by ascending value, one for each column that you want to index. In this way, each time a query is made with a constraint of type\cod{WHERE x > c}, a binary search is made in the index to reduce the search space.

From here things were looking nice. There was only a major inconvenient, and that was that ALL the information should reside in memory. Each time the product started, it had to load the data of a CSV file into the memory and generate the indexes. This did not take too long but it was annoying.

Shortly before the first release of the product all this collapsed. Basically because for the release, information was generated for ALL the social graph of a country, which included more than 100 million records, so the information no longer fitted into memory.

Again pressing the panic button, the solution arose: have the data saved on disk. The data was saved in ``raw'' mode on disk, one file per column, so that you could make mmap of each file to load the data of each column, with the additional advantage that the operating system is responsible for caching everything automatically .

And this is roughly how VTor was born.

But as it usually happens, the requirements started to grow, and to grow:
\begin{itemize}
\item We need to make histograms.
\item We need to filter the outliers of the histograms, so we need support to calculate percentiles.
\item We need support for String columns, even though we had sworn at first that we were not going to need.
\item We need a\cod{COUNT (DISTINCT)}.
\item We need to be able to join tables!
\item Can we optimize such and such graph?
\item Why does not VTor support the OR comparison?
\item Can we add NULLs?
\end{itemize}

So what we predicted to be a few hundred lines of code began to grow and grow. Remember that at this point all the queries were hard-coded, so any new feature that we wanted to include was a lot of work since the architecture was not intended for anything more than a very punctual pair of use cases. Every time we had to add something, we suffered.

%!TEX root = snel.tex

%-----------------------------------------------------
\subsection{The Birth of Snel}
%-----------------------------------------------------

In addition to the rigidity of VTor, there were two things that were very bothersome:
\begin{enumerate}
\item It was done in C. Segmentation faults were occurring every day.

\item There were many macros everywhere to handle different types of data in the same way. For example, it is not the same to solve something that executes\cod{SUM (column_int32)} than something which executes\cod{SUM (column_int64)}. 
This not only causes problems when programming because you have to duplicate a lot of code, but it also had an impact on execution time, since for each value that is generated you have to call the function with the corresponding data type. Basically the same as a virtual table in C++ (not to be confused with the ``Virtual Tables'' of SQLite that are a completely different thing).

\end{enumerate}

It was at this point that we learned how a database engine really works. It turns out that what you learn in college on relational algebra and those kind of things are not an invention and can be used in real life!

We studied the paper ``Efficiently Compiling Efficient Query Plans for Modern Hardware''~\cite{neumann2011efficiently}. The main problem, they say, is the dispatch of methods in run time. The same problem we mentioned earlier with the data types! This causes thousands or millions of virtual function calls in the execution of a query plan, which breaks the execution pipeline of the CPU.

The solution proposed by this paper was to compile the queries to machine code using LLVM~\cite{lattner2004llvm}, which solves the problem of virtual functions by directly generating the necessary code for each data type.
As an experiment, we started to implement the ideas of this paper. The main ideas of \cite{neumann2011efficiently} are:
\begin{enumerate}
\item Processing is data centric and not operator centric.
Data is processed such that we can keep it in CPU
registers as long as possible. Operator boundaries are
blurred to achieve this goal.
\item Data is not pulled by operators but pushed towards
the operators. This results in much better code and
data locality.
\item Queries are compiled into native machine code using
the optimizing LLVM compiler framework.
\end{enumerate}
An additional advantage was that using relational algebra to solve queries gave a lot more flexibility to solve queries than VTor did. So, we went on developing and developing.

It took one year but in the end we were able to finish it. We called this new engine ``SNEL'' as an acronym for ``SQL Native Execution for LLVM''. Well, that's the excuse, actually we chose Snel because that means ``fast'' in Dutch and the name seemed right to us.

When developing Snel we made the following decisions:

\begin{enumerate}
\item We were going to continue using the mechanism of SQLite virtual tables, which was already implemented.

\item We were going to use the same VTor tables. So only the switch of the virtual table module from VTor to Snel could be made and the results had to be the same, with an improvement in performance. Therefore, we could continue using the mechanism of mmaping the files with the data of the columns.

\item We did it in C++, since we were hating C and also the LLVM API is in C++. While C++ brought a few problems, the number of segmentation faults and memory leaks was greatly reduced. Now at least we had exceptions!

\end{enumerate}

% Section 2
%!TEX root = snel.tex

\section{The Architecture of Snel}

Broadly speaking, Snel has 3 important parts:

\begin{enumerate}
\item The storage system.

\item The SQL translator to relational algebra and optimizer.

\item The translator from relational algebra to IR. The Intermediate Representation (IR) is a low-level programming language similar to assembly, and can be considered as the ``bytecode'' of LLVM.

\end{enumerate}

\subsection{Storage System}

Snel stores the tables in a very simple way: each column is stored in a single file, in ``flat'' format. That is, if for example we have an int32 column, the file will have 4 bytes for each value. This allows you to map the file to memory and use it as if it were an array directly, without any intermediate conversion. The big problem with this is that it wastes a lot of space, since the vast majority of data are NULLs or 0 and could easily be compressed. Each column file has a\cod{.snelcol} extension.

The indexes are stored in a similar way to the columns, except that each value of the index is a tuple (value, rowid). The rowid is 64 bits, so each index entry is (sizeof (value) + 8) bytes. This also wastes a lot of space. Each index file has a\cod{.snelidx} extension.

In addition to the column and index files, there is a file with extension\cod{.snel} that is nothing more than a JSON with the description of the schema of the table: what type of data each column has, if they are indexed or not, if NULLs are allowed, etc.

\subsubsection*{Text Columns}

What we said earlier makes sense for columns with fixed-length data types: integers and floating point. However, the text columns have a special format.

Each\cod{.snelcol} file that stores a column of text has the following format:
\bbash
String 1\0
String 2\0
...
String n\0
Sync bytes ('SB')
Offset to String 1 (-1 if NULL)
Offset to String 2 (-1 if NULL)
...
Offset to String n (-1 if NULL)
Offset to the offset of String 1 (yo dawg)
\end{lstlisting}

At first glance, this last value does not seem to make sense. However, it is necessary to know where the list of offsets begins within the file quickly.
The indices of the text columns do not include the value, they are only a list of rowids ordered according to the string value from smallest to greatest. That is, it is not a COVERING INDEX.

\subsection{Translator and Optimizer}

True to its VTor roots, Snel translates a SQL query provided by SQLite to a relational algebra tree. The SQL query comes as a tree of tokens and the module\cod{snel_sqlite} (which is the interface between SQLite and Snel) translates that into a structure of queries understood by Snel.

This structure is quite simple. Viewing the\cod{src/query.hpp} file you can see what SQL features Snel supports:
\medskip
\begin{lstlisting}[style=CStyle]
boost::ptr_vector<SQLExpression> fields;
ConstraintTree constraintTree;
std::set<const Table*> sourceTables;
UniqueExpressionList groupBy;
std::vector<OrderClause> orderBy;
uint64_t limit = NO_LIMIT;
uint64_t offset = 0;
bool distinct = false;
\end{lstlisting}

\noindent
The structure is:
\begin{description}
\item[fields:] corresponds to the SELECT clause of SQL. It contains the list of expressions that are going to be solved (ex: col1, SUM (col2), COUNT (*), etc.)
\item[constraintTree:] corresponds to the WHERE clause of SQL. Example: (col1> 10 AND (col2 <20 OR col3 = 5)). It is represented by a tree where each node is a Boolean operator.
\item[sourceTables:] corresponds to the FROM clause of SQL, that is, the list of tables that will be queried.
\item[groupBy:] expressions by which GROUP BY is going to be done.
\item[orderBy:] as in SQL.
\item[limit:] as in SQL.
\item[offset:] as in SQL.
\item[distinct:] flag that indicates whether the query is DISTINCT or not, that is, whether it will return duplicate values or not.
\end{description}

This is basically the subset of SQL that Snel understands. Note that it does not support subqueries, at least not directly, although it would not be terribly difficult to add them.

Snel's translator will grab this structure and turn it into a tree of relational algebra operators. Without going into too much detail (that comes later), Snel converts the following query:
\begin{lstlisting}[style=SQLStyle]
SELECT foo, COUNT(*) FROM table1 WHERE bar > 2 GROUP BY foo 
	ORDER BY foo DESC LIMIT 10 OFFSET 2
\end{lstlisting}

The optimizer will grab this query plan and apply several optimization steps to make the query run faster, such as verifying whether the table has indexes to hit only a subset of the rows.

\subsection{Translator to IR}

Once the optimized query plan is generated, it is translated into IR code, which in turn is compiled into machine code by LLVM and then the query is executed. I will explain this in more detail later.

%!TEX root = snel.tex

\subsection{The Snel Code and Class Hierarchy}

The Snel code is divided into three parts:
\begin{enumerate}
\item The main library,\cod{libsnel.so}. It is the dynamic library to which the other two parts link. The code is inside the\cod{src/} directory.
\item The command line. It allows to do several tasks related to the maintenance of tables (creation / import, generation of indexes, list contents, etc). Located in the directory\cod{cli/}.
\item The interface with SQLite. As mentioned before, SQLite allows you to incorporate external engines, and this module allows queries to Snel tables from SQLite directly. The code is in the\cod{snel_sqlite/} directory.

\end{enumerate}

Apart from these three main parts, we can also find:
\begin{enumerate}
\item Testing code in the\cod{tests/} directory.
\item A utility to generate Snel tables from Spark in the\cod{spark-snel-connector/} directory.
\item A script that allows you to add columns to Snel tables in the\cod{scripts/} directory.

\end{enumerate}

% \subsection{The Main Library: Class Hierarchy}

We now describe the class hierarchy of the main library.
% First of all, I apologize for the lack of documentation of the classes. My excuse is nothing original: most of the code was experimental and was growing organically, without planning.
The organization of the Snel code into classes corresponds to the architecture previously described and has three main parts: storage, relational algebra engine and IR translator.

The classes of the library are described in the following sections.

\subsubsection{Storage}

\begin{description}

\item[BaseVector:] Represents a vector of fixed-size elements. It is similar to a\cod{std::vector<>} but with the difference that the size of the elements is defined in runtime. This allows us to generate vectors from the JIT code.
A vector has an associated allocator (see below), which allows the vector to reside in memory or to be saved to disk in a transparent form.

\item[Allocator:] It is the interface for vector allocators. There are two types of allocators.
\begin{description}
\item[MemoryAllocator:] For vectors that are going to be generated in RAM.
\item[MMapAllocator:] For vectors that have a file in the filesystem as backup.
\end{description}

\item[BaseColumn:] It is an interface that represents a column of a table. An interface is used since afterwards there will be a subclass for each type of column (int [8, 16, 32, 64], float and string).
\begin{description}
\item[Column<T> and StringColumn:] specific subclasses for each type of column.
\end{description}

\item[BaseIndex:] Interface that represents an index associated with a column. Note that this class descends from BaseVector, since all the indexes are vectors with elements of fixed size.

\item[Table:] Table with its columns.

\end{description}

\subsubsection{SQL / Relational Algebra}

\begin{description}

\item[Query:] Structure that represents a SQL query.

\item[QueryPlan:] Represents the relational algebra tree of a query plan and contains the code to generates the query plan from a Query instance.

\item[QueryPlanOptimizer:] Takes a QueryPlan, applies several optimization steps, and returns another QueryPlan, theoretically more optimal. This class is VERY complex. As I read once: ``Query optimization is not rocket science. When you flunk out of query optimization, we make you go build rockets.''

% \begin{description}
\item[QueryPlanParallelizer:] This class is another step of optimization, but I put it in a separate file because it is a bit more complicated than the others. It analyzes the tree of a query plan and decides which parts can be parallelized to execute them in several threads.

% \end{description}

\item[*-Expression:] All classes that end in -Expression that are in the subdirectory\cod{expressions/} represent SQL expressions. For example, the function\cod{SUM(column1)} is a SUM expression composed with a column expression.

\end{description}

\subsubsection{IR Translator}

\begin{description}
\item[BaseBuilder:] It is a subclass of IRBuilder of LLVM that has some useful extensions.

% \begin{description}
\item[QueryPlanBuilder:] IR Builder that has some specific utilities to generate IR code from query plans.
% \end{description}

\item[LLVMWrapper:] Utilities that allow to call C++ functions from JIT code. For now it is only a wrapper for the methods of the BaseVector class.

\item[QueryContext:] A JIT execution context is simply a set of variables that are visible by an execution thread. When a query is divided into several threads, each thread will have its own execution context (all threads will have the same variables but with different values).

\item[*-Operator:] The classes that end in -Operator that are in the\cod{operators/} subdirectory represent the different relational algebra operators supported by Snel. Later each operator is explained in detail.
\end{description}

%\subsubsection{Miscellaneous}
%
%\begin{description}
%
%\item[Benchmark:] Utility that allows benchmarks of code sections.
%\item[QueryPlanTreeGraph:] Generates DOT files from a query plan to be able to visualize them. These files can be viewed with the xdot utility.
%\item[Datatype:] Interface that represents the different types of data supported by Snel.
%
%\end{description}

% Section 3
%!TEX root = snel.tex

\section{Implementation of Snel}

Here we will explain in as much detail as possible how Snel is implemented. If you have to get your hands dirty and modify something, this is a good place to begin to understand how everything works and why things are made the way they are made.

\subsection{First of All: How to Compile}

First of all, it is necessary to install LLVM. The latest supported version is 3.6.2, since the API from 3.7 is not backwards compatible. At some point you would have to change the code to update it to the new API.

Since Ubuntu as of 16.04 comes with LLVM 3.8 installed by default, we will have to compile and install LLVM in a separate directory. The steps to download, compile and install LLVM in\bash{/opt/llvm-3.6.2} are:

\bbash
$ wget http://llvm.org/releases/3.6.2/llvm-3.6.2.src.tar.xz
$ tar xf llvm-3.6.2.src.tar.xz && cd llvm-3.6.2.src
$ mkdir build && cd build
$ ../configure --enable-jit --enable-optimized --enable-targets=host 
	--enable-assertions --prefix=/opt/llvm-3.6.2
$ make -j3 && sudo make install
\end{lstlisting}

This takes a long time, so go get a coffee.
Once LLVM is installed, you have to download the dependencies:
\medskip
\begin{lstlisting}[style=BashStyle]
$ sudo apt-get install cmake g++ libgomp1 libboost-all-dev libreadline-dev 
	zlib1g-dev mawk liblog4cxx-dev libyaml-cpp-dev
\end{lstlisting}

Now you have to compile the famous sqlite3-grandata. Assuming that the repository checkout is in 
\medskip
\lstinline[style=BashStyle]{~/src/dpi}:
\begin{lstlisting}[style=BashStyle]
$ cd ~/src/dpi/sqlite3-grandata
$ mkdir -p build/release && cd build/release
$ cmake ../.. -DCMAKE_BUILD_TYPE=Release
$ make -j3 && sudo make install
$ # Verify that everything works fine
$ sqlite3-grandata --version
3.7.17 grandata
\end{lstlisting}

Now we proceed to compile Snel. We're going to do it in Debug mode to be able to... well, debug:
\medskip
\begin{lstlisting}[style=BashStyle]
$ cd ~/src/dpi/projects/snel
$ mkdir -p build/debug && cd build/debug
$ cmake ../.. -DCMAKE_BUILD_TYPE=Debug 
	-DLLVM_DIR=/opt/llvm-3.6.2/share/llvm/cmake
\end{lstlisting}

\subsection{Testing}

To check that everything is working well, we can run the tests. They are written using the Google testing framework and can be run together using CTest, a test running tool that comes with CMake:

\medskip
\begin{lstlisting}[style=NoStyle]
$ cd $PATH_TO_SNEL/build/debug/test
$ ctest
Test project /home/marcelo/dpi/projects/snel/build/debug/tests
    Start 1: test_basics
1/6 Test #1: test_basics ......................   Passed    0.09 sec
    Start 2: test_single_table
2/6 Test #2: test_single_table ................   Passed    0.37 sec
    Start 3: test_null
3/6 Test #3: test_null ........................   Passed    0.29 sec
    Start 4: regression_tests
4/6 Test #4: regression_tests .................   Passed    1.22 sec
    Start 5: test_config
5/6 Test #5: test_config ......................   Passed   39.21 sec
    Start 6: test_sql
6/6 Test #6: test_sql .........................   Passed    5.28 sec

100% tests passed, 0 tests failed out of 6

Total Test time (real) =  46.48 sec
\end{lstlisting}

%!TEX root = snel.tex

\subsection{Creating Snel Tables}

At the moment Snel tables can not be created using the SQL interface but must be created ``from the outside''. To create the tables you must use the command\bash{snel import}, which will import data in plain text from some source to a new Snel table. Let's see:
\medskip
\begin{lstlisting}[style=NoStyle]
$ snel import --help
Usage: snel import [options] <table name> <schema file> 
	<output directory> [input file]
Options:
   -h              This help message
   -s <separator>  Field separator character (default: "|")
   -b <size>       Buffer size, in number of rows (default: 100,000)
   --safe          Check for invalid input (SLOWER)
   --null-repr     Representation for null fields
   -v              Be verbose
\end{lstlisting}

We then need two things: a data source (usually a CSV file) and a file to describe the schema of the table to be created.
The schema file has several lines with the following format:
\medskip
\begin{lstlisting}[style=NoStyle]
<column name 1> <tipo> [NULLABLE] [INDEXED],
<column name 2> <tipo> [NULLABLE] [INDEXED],
...
<column name n> <tipo> [NULLABLE] [INDEXED]
\end{lstlisting}
Where the type of the column can be one of the following:
\begin{itemize}
\item BOOLEAN, BOOL or BIT: value 0 or 1
\item INT8 or CHAR: 8-bit integer
\item INT16 or SHORT: 16 bit integer
\item INT32 or INT: 32-bit integer
\item INT64 or LONG: 64-bit integer
\item FLOAT: 32-bit floating point (Note: doubles are not supported)
\item STRING or TEXT: variable length text
\end{itemize}

By default the columns DO NOT accept NULLs, unless the NULLABLE flag is specified. In the same way, the columns are not indexed unless the flag INDEXED is specified.
By convention, the table schema specification files are saved in a file with the\bash{.snelschema} extension.
Going back to the\bash{snel import} command, the most common use case is to grab a file with extension\bash{.csv.gz} and to pipe it to that command. Something like this:
\bbash
$ # If you do not have the pv command installed, I recommend it STRONGLY. 
$ # You will thank me.
$ # In this case we use the tail -n +2 to remove the headers from the csv file
$ pv archivo.csv.gz | zcat | tail -n +2 | 
	snel import table_name schema.snelschema /tmp/sarasa
\end{lstlisting}

If all goes well, you will see in\bash{/tmp/sarasa} many files with the name
\bbash
table_name-<column_name>.snel[col|idx]
\end{lstlisting} 
and a file named\bash{table_name.snel} which is the JSON that contains the data and metadata of the table. 
These files contain the raw data of the columns. So much so that if you read the files to an array of C and you cast it to the correct data type, you will be able to read all the values of the corresponding column (except for the columns of type text, whose format I already explained before).

\subsubsection{Update values or add columns to a Snel table}

In its origins, both VTor and Snel were read-only. But since the format is so simple nothing prevents us from updating the values directly or adding new columns to an existing table.
The\bash{snel merge} command allows you to ``merge'' two tables using a column as a merge key. Let's see a typical example: we have the column clients where the primary key is the MDN column, and we want to add a new variable to the table. Suppose that this variable is called\bash{eye_color} and is a number between 0 and 3. Suppose we also want to update the\bash{height} variable that already existed in the customer table but we did a super algorithm that inferred this variable with an accuracy of +/- 1 cm in comparison with the existing variable that has a precision of +/- 50 cm (yes, a crap).
Usually these two variables come in a CSV that looks like this:

\bbash
mdn|eye_color|height
B0A00F35B1F5DEAD84DB2D28BD883094|0|1.77
9386711E612687A86DF876B7D76FB514|2|1.52
3328EFF2A02D215A16F52F46A83CAE1B|3|1.85
\end{lstlisting}

And suppose we only have information for these three MDNs. The problem is that we can not create a table and put the new column ``by hand'' by copying the\bash{.snelcol} file because nothing would indicate that the MDNs are in the same order.
The solution is then to create a new table with this information using the\bash{snel import} command:
\bbash
$ cat new_variables.csv | tail -n +2 | snel import new_variables
	new_variables.snelschema /tmp/new_variables
\end{lstlisting}
Where the file\bash{new_variables.snelschema} is:
\bbash
mdn STRING INDEXED,
eye_color INT8 NULLABLE,
height FLOAT NULLABLE
\end{lstlisting}
This will create\bash{/tmp/new_variables/new_variables.snel} and all the associated column files.
Now let's see the\bash{snel merge} command:
\bbash
$ snel merge --help
Usage: snel merge [options] <source table> <dest table> <key column name>
Options:
   -h              This help message
\end{lstlisting}

This will merge the source table\bash{new_variables} in the target table (clients) using the key column (mdn).
The command for the merge is then:
\bbash
$ snel merge /tmp/new_variables/new_variables.snel /path/to/clients.snel mdn
\end{lstlisting}

This will do the following:
\begin{enumerate}
\item It will create the column\bash{eye_color} in the table clients and will set the values in the corresponding rows. The rest of the rows will be NULL.
\item It will update the corresponding values of the height column.

\end{enumerate}

In order to simplify all these steps, there is a script in\bash{scripts/snel-merge-csv.py} that automates all this. In this case it would be executed in the following way:

\bbash
$ cat new_variables.csv | tail -n +2 | 
	scripts/snel-merge-csv.py -c eye_color:int8 -c height:float 
	/path/to/clients.snel mdn
\end{lstlisting}

This will create a temporary table in\bash{/tmp} (make sure there is enough space!) And it will do the merge automatically.

Note that you can't append data to a Snel table. There is nothing to prevent it, it's just that that functionality is not implemented.

%!TEX root = snel.tex

\subsubsection{Generating tables with Spark}

In the Snel repository there is a connector for Spark
that allows generating Snel tables using Spark. This involves two steps:
\begin{enumerate}
\item Generation of partitioned tables in HDFS (Hadoop Distributed File System).
\item Merging the partitions from HDFS to where the table will be used.
\end{enumerate}

So, how is a Snel table generated from Spark? There is an implicit function called\cod{saveToSnel} that applies to Resilient Distributed Datasets (RDD) of type\cod{RDD[Seq[Option[Any]]]}. The function looks like this:
\bbash
def saveToSnel(destPath: String, snelSchema: SnelSchema, 
	compress: Boolean=true): Unit
\end{lstlisting}

\begin{itemize}
\item	destPath: Path where the table will be saved. For example:\bash{hdfs://host_hdfs.com:1234/user/pepe/tabla_snel}

\item	snelSchema: The schema of the table. It is basically a table name and an array with the specification of each column. The specification of each column is an instance of one of the following classes:
\begin{itemize}
\item BoolColumn
\item Int8Column
\item Int16Column
\item Int32Column
\item Int64Column
\item FloatColumn
\item TextColumn
\end{itemize}

\item	compress: Indicates whether the data will be compressed on HDFS. Unless there is a really strong motive, you usually have to leave this parameter as true.

\end{itemize}

Now, we must bear in mind that this is very nice when it comes to making type conversions. Each value within the RDD is a Seq that represents the values of the row, and there must be as many values as the number of columns. Each value can be NULL or non-NULL, that is why an Option is used. If the value is non-null, the value is cast to the correct data type, for example if the schema says that this value corresponds to an int16, it will do\cod{value.asInstanceOf[Short]}, and if this fails it will throw an exception.

So, in order to make the conversion, given a generic RDD, it is necessary to do the following steps:
\begin{enumerate}
\item Generate the Snel schema, doing something like this:
\bcod 
val snelSchema = SnelSchema(tableName = "una_tabla", 
   columns = Array(
   Int16Column("una_columna_int16", isNullable=true, isIndexable=false),
   FloatColumn("una_columna_float"),
   ... etc ...
))
\end{lstlisting}

\item Convert all RDD values to an\cod{Option[Any]} with the correct data type. This is the most expensive step. The mapping between types of columns and values is as follows:
\bcod
Type of column Type of Scala
BoolColumn	Boolean
Int8Column	Byte
Int16Column	Short
Int32Column	Int
Int64Column	Long
FloatColumn	Float
TextColumn	String
\end{lstlisting}

To do this, you can obviously map the RDD so that it has the desired format.

\item Once the RDD is generated with the desired format, call\cod{rdd.saveToSnel (...)} with the appropriate parameters. Note that for this function to be available, we must first do\cod{import com.grandata.snel.spark.connector._}
so that the Scala compiler puts the function as implicit.
\end{enumerate}

This will trigger a Spark task that will generate the table in HDFS. When viewing the list of generated files, you can see the following:
\bno
bash-4.1# bin/hdfs dfs -ls /user/marcelo/test_table
Found 5 items
-rw-r--r--   3 [...] /user/.../test_table/_SUCCESS
drwxr-xr-x   - [...] /user/.../test_table/test_table-floatcol.snelcol
drwxr-xr-x   - [...] /user/.../test_table/test_table-int32col.snelcol
drwxr-xr-x   - [...] /user/.../test_table/test_table-textcol.snelcol
-rw-r--r--   3 [...] /user/.../test_table/test_table.snel

bash-4.1# bin/hdfs dfs -ls /user/marcelo/test_table/test_table-floatcol.snelcol
Found 3 items
-rw-r--r--   3 [...] /user/.../test_table/test_table-floatcol.snelcol/part-r-00000.gz
-rw-r--r--   3 [...] /user/.../test_table/test_table-floatcol.snelcol/part-r-00001.gz
-rw-r--r--   3 [...] /user/.../test_table/test_table-floatcol.snelcol/part-r-00002.gz
\end{lstlisting}

As you can see, it creates a directory for each column with the parts of the columns inside.
Snel can not access this information directly, so you have to run another Spark job to download these files and assemble them in the place where they will be used. For this, you have to call the function\cod{com.grandata.snel.HDFSMerge.mergeTable()}.

I strongly recommend seeing the tests in this package to see how everything works, mainly the test of the HDFSMergeTest class, which is an integral test.

%!TEX root = snel.tex

\subsection{Executing Queries}

For now, the only way to execute SQL queries on one or more Snel tables is by using the SQLite interface, which translates SQL queries into the structure that represents a query in Snel.

The SQLite interface for Snel is a dynamic library\cod{libsnel-sqlite.so} that should load automatically when loading the binary\cod{sqlite3-grandata}. If this does not happen, you have to specify the path by hand with the command\cod{.load}. For example, when Snel is compiled in debug mode, I do not want to install it in\cod{/usr/local} so I load it directly from where it is generated:
\bno
$ sqlite3-grandata 
SQLite version 3.7.17 grandata
VTor version vtor-1.6.7-release
Enter ".help" for instructions
Enter SQL statements terminated with a ";"
sqlite> .load /home/.../snel/build/debug/snel_sqlite/libsnel_sqlited.so 
Initializing Snel 0.9.1-debug (git: develop/8906b0f)...
[DEBUG] [0x7f53ad3ad7c0] ... snel - Initializing LLVM native target
[DEBUG] [0x7f53ad3ad7c0] ... snel - Finished initializing
[DEBUG] [0x7f53ad3ad7c0] ... snel - Setting locale to en_US.UTF-8
sqlite>
\end{lstlisting}

Once the module is loaded in SQLite, we must indicate that we want to create an external or virtual table using the Snel module:
\bno
sqlite> CREATE VIRTUAL TABLE tabla USING SNEL('/path/a/tabla.snel');
sqlite> SELECT COUNT(*) FROM tabla;
123456
\end{lstlisting}

Here all the magic happened: the query was translated into relational algebra, the resulting tree was translated into IR code, that code was compiled into machine code and that code was executed.

\subsubsection*{Types of queries supported}

Snel supports the following SQL functionality:
\begin{itemize}
\item SELECT of columns and aggregated columns.
\item Aggregation functions: \\
	\cod{SUM, AVG, MAX / MIN, COUNT, COUNT (DISTINCT)}.
\item Aggregation functions for histograms: \\
	\cod{BIN, BIN_MIN, BIN_MAX}. \\
	These are unique features of Snel.
\item WHERE clauses:
\begin{itemize}
\item <column> <op> <expr>: where <op> can be >, >=, <, <=, = or !=, and <expr> can be a constant or the name of another column.
\item Boolean operators:\cod{AND, OR, NOT}.
\item Clause <column> IN (<subquery>) where <subquery> is a query supported by Snel. This is the only type of subqueries that Snel supports for now.
\end{itemize}

\item GROUP BY.
\item ORDER BY.

\end{itemize}

This is an example of a complex query supported by Snel:
\bsql
SELECT BIN(latitude, 10) AS bin_lat, BIN(longitude, 10) AS bin_long,
   BIN_MIN(latitude), BIN_MAX(latitude), BIN_MIN(longitude),
   BIN_MAX(longitude), COUNT(*), SUM(col1), AVG(col2), 
   COUNT(DISTINCT col3) AS cnt
FROM table
WHERE 
   (latitude BETWEEN -10 AND 10
   AND longitude BETWEEN -20 AND 20
   AND age > 30)
   OR age IN (SELECT ages FROM otra_tabla)
GROUP BY bin_lat, bin_long
ORDER BY cnt DESC
LIMIT 10
\end{lstlisting}

%\subsubsection{Unsupported Queries}
%
%\begin{itemize}
%\item Complex expressions. For example: \cod{SELECT column + 1 FROM table}
%\item Subqueries: \cod{SELECT x FROM (SELECT x FROM ...) AS s}
%
%\end{itemize}

% Section 4
%!TEX root = snel.tex

\section{The Life of a Query}

Let's see what happens between writing a query and returning its result. Broadly speaking, the following steps are executed:
\begin{enumerate}
\item The SQLite parser converts the string with the SQL query to an AST tree (Abstract Syntax Tree).
\item SQLite "asks" the Snel module if it is able to resolve that query -- performed by\cod{sqlite3_module.cpp}, function\cod{snel_xCanOptimizeSelect}. Snel analyzes the query and, if the query has a supported structure, returns true.
\item Let's assume that the query can be executed by Snel. In that case, call the\cod{snel_xExecuteCustomSelect} function that gives control of the execution to Snel.
\item Snel converts the AST to a relational algebra tree, the query plan. This is an almost direct translation. After doing this, it applies several optimization steps to the query plan to have an optimized query plan (see the Query::executeQuery function).
\item The query plan is translated into LLVM IR code. See function QueryPlanBuilder::generateTopLevelQueryPlanIR.
\item The LLVM execution engine is created with the aggressive optimization parameter so that the queries are fast.
\item The IR of the query plan is compiled to machine code using the execution engine. The result of the compilation is saved in an object of type ExecutedQuery, which is the one that will be the owner of the LLVM module and will provide a cursor to iterate the resulting rows.
\item The control is returned to SQLite. SQLite will then call the\cod{snel_xNext} function that will advance the cursor row by row and\cod{snel_xColumn} 
that will return the values of the resulting rows.
\item When\cod{snel_xNext} indicates EOF, SQLite calls\cod{snel_xClose}, which frees all resources.
We will see these steps in more detail, with references to the code to make it easier to understand.
\end{enumerate}

\subsection{SQL Translator}

The first step is done by SQLite and is simply the translation of the SQL query to an AST. This is a standard compiler process.
Next, SQLite invokes the\cod{snel_xCanOptimizeSelect} function, passing the AST as a parameter. The AST is an instance of the \emph{Select} type of SQLite, which looks like this:

\bcod
struct Select {
  ExprList *pEList;      /* The fields of the result */
  u8 op;                 /* One of: TK_UNION TK_ALL TK_INTERSECT TK_EXCEPT */
  u16 selFlags;          /* Various SF_* values */
  int iLimit, iOffset;   /* Memory registers holding LIMIT & OFFSET counters */
  int addrOpenEphm[3];   /* OP_OpenEphem opcodes related to this select */
  double nSelectRow;     /* Estimated number of result rows */
  SrcList *pSrc;         /* The FROM clause */
  Expr *pWhere;          /* The WHERE clause */
  ExprList *pGroupBy;    /* The GROUP BY clause */
  Expr *pHaving;         /* The HAVING clause */
  ExprList *pOrderBy;    /* The ORDER BY clause */
  Select *pPrior;        /* Prior select in a compound select statement */
  Select *pNext;         /* Next select to the left in a compound */
  Select *pRightmost;    /* Right-most select in a compound select statement */
  Expr *pLimit;          /* LIMIT expression. NULL means not used. */
  Expr *pOffset;         /* OFFSET expression. NULL means not used. */
};
\end{lstlisting}

As we can see, it is not a science. Each SQLite expression (Expr) contains several details about the types of SQL expressions, for example, if it is a constant, if it refers to a column, if it is an aggregation function, if it is a subquery, etc.
Snel will try to translate this structure into a structure of the type snel::Query.
This translation is done by traversing the AST recursively, section by section (SELECT fields, WHERE, GROUP BY, etc). At any time the translator may find an unsupported SQL feature such as subqueries, in that case the translation process ends with an exception.
Even if the translation is successful, it may happen that the query for some reason is not supported by Snel. In this case, SQLite is informed that the query can not be processed internally by the engine, thereby giving control of the execution of the query to SQLite.
If the query can be translated into a Snel query and is supported, the\cod{xCanOptimizeSelect} function returns true. This tells SQLite that the execution of the query will be handled internally by Snel, which as explained before is a much faster mechanism.

\subsection{Execution of the Query}

If the query can be executed by Snel, SQLite will yield the query execution by calling the\cod{xExecuteCustomSelect} function, where the AST of the query will be passed again and Snel will do the same translation as before.
Once the translation of the AST is done, the function snel::Query::executeQuery is called, where ALL the magic happens: translation to relational algebra, optimization and compilation to machine code.
The result of this whole process is a cursor: basically an object that can be advanced with the function next() and from where the individual values of the columns can be brought.

\subsection{Query to Relational Algebra}

An object of the Query type is translated into a tree of relational algebra operators. The operators implemented in Snel are:
\begin{description}
\item \cod{ACCUMULATE}: It is an operator that simply stores the rows in memory for further processing by another operator. It is useful for example when we want to buffer the result of a sub-tree.
\item \cod{AGGREGATOR}: It is the operator that is responsible for solving aggregation expressions such as SUM, COUNT, AVG, etc.
\item \cod{BINNER}: It is a special operator of Snel. It is used to group values in different bins, generally used to calculate histograms of values.
\item \cod{CONSTRAINT}: It is responsible for filtering the columns that do not satisfy the given constraints, for example "A = 1 OR (B> 2 AND c <3)".
\item \cod{DEBUG}: Useful operator to debug, the only thing it does is to show the rows that pass through this operator by the standard output.
\item \cod{LIMIT}: Limit the number of rows to a given value.
\item \cod{MERGEJOIN}: Implements a sort-merge-join between tables. For now this is the only algorithm implemented, it would be good to implement a hash-join in the future.
\item \cod{PARALLEL_AGGREGATOR}: Same as the aggregator but allows the aggregations to run in parallel using several CPUs, and is responsible for joining the results.
\item \cod{SORT}: Sorts the rows that arrive according to some criteria (usually the value of a given column).
\item \cod{FULL SCAN}: Full scan of a table. Run through all the columns sequentially.
\item \cod{INDEX SCAN}: Scan a table through an index. The table is traversed non-sequentially.
\item \cod{XJOIN} (or CROSS JOIN): Makes an ``all against all'' join, and is $O (n^2)$ (Cartesian product).
\end{description}

In a first step, the translation of the query to AR is quite direct. For example, a query like 

\makebox[\linewidth]{\cod{SELECT a, COUNT (*) FROM table WHERE c <2 GROUP BY a}  } 

is translated simply as 

\makebox[\linewidth]{\cod{FULL SCAN -> CONSTRAINT -> AGGREGATE} . }

The second step is to grab the AR tree and apply several optimization steps. The steps that are applied are (in order):
\begin{enumerate}
\item Optimization of JOINs: It evaluates whether the CROSS JOINS can be replaced by a MERGE JOIN. This occurs if there is a constraint of type table1.column1 = table2.column2 and both columns are indexed.
\item Relocation of CONSTRAINT: This step tries to "lower" the constraints that apply to a single table immediately after the SCAN associated with that table. This serves two purposes: to quickly filter the rows that do not satisfy the constraints and to apply subsequent optimizations. In the literature this is known as predicate pushdown.
\item Indexing: It evaluates whether there is a FULL SCAN followed by a CONSTRAINT. If so, it checks whether an index can be applied, and if it can, it replaces the FULL SCAN with an INDEX SCAN (the CONSTRAINT is not replaced since it may have other constraints that do not apply to the index).
\item SORT Relocation: Try to pushdown the SORT operators. This step is necessary for the next one.
\item SORT Elimination: Check whether some SORT have become obsolete. The SORT is a costly operation and should be avoided as much as possible.
\item Parallelization: Finally, it checks whether the query can be parallelized. There are some conditions that a query must fulfill in order to be parallelized (basically, all the aggregation operators are parallelizable). If it is parallelizable, a new operator is added as the root of the query, the\cod{PARALLEL_AGGREGATOR}.
\end{enumerate}

Let's see a simple example query:
\bsql
SELECT COUNT(*) FROM table1 WHERE int8col1 > 3
\end{lstlisting}

Let's see how this query is translated into relational algebra. Opening an existing database, first disable multithread execution to simplify things, and let's see the translation:
\bsql
sqlite> SELECT SNEL_SET_MAX_THREADS(1);   --Turns off multithreading
sqlite> SELECT COUNT(*) FROM table1 WHERE int8col1 > 3;
48429032
sqlite> EXPLAIN QUERY PLAN SELECT COUNT(*) FROM table1;
0|0|0|AGGREGATE (cost: 100000000) { CONSTRAINT [table1.int8col1 > 3] 
	{ FULL SCAN FOR TABLE 'table1' (100000000 rows) } }
\end{lstlisting}

The translation to AR is then
\bbash
AGGREGATE (cost: 100000000) { CONSTRAINT [table1.int8col1 > 3] 
	{ FULL SCAN FOR TABLE 'table1' (100000000 rows) } }
\end{lstlisting} 
This is read "from the inside out": first make a FULL SCAN of the table, filter by the specified column and then AGGREGATE the rows. The aggregation in this case is simply to count the number of rows that satisfy the constraint. In pseudocode:
\bcod
num_rows = 0
while table1.nextRow():      # FULL SCAN
  if table1["int8col1"] > 3: # CONSTRAINT
      num_rows++             # AGGREGATE
\end{lstlisting}

When executing the previous query, the IR code will be stored in a file called\cod{queryplan.ll} and the AR tree will be saved in the\cod{queryplan.dot} file. 
Running xdot in this last file, we can generate a graphic representation of the query.

\subsection{Translation to IR Code}

Note that in order to understand this section, it is important to understand a little how LLVM works. I recommend following the Kaleidoscope tutorial to understand the API.
Broadly speaking, we can say that the translation is ``direct'' in the sense that each relational algebra node is translated to IR independently of the others.
Each AR tree is compiled to IR code inside a module that looks like this:

\bcod
; ModuleID = 'queryplan'

%TopLevel.ContextType = type { ... }
%TopLevel.ResultRow = type <{ ... }>

@contextInstance = global %TopLevel.ContextType zeroinitializer

@output.0 = global algun_tipo 0
@output.1 = global algun_tipo 0
...
@output.n = global algun_tipo 0

define private void @initializeContext_TopLevel.Context
	(%TopLevel.ContextType* %contextPtr) {
  ...
}

define void @queryPlanInit(%TopLevel.ContextType* %contextInstancePtr) {
  ... 
}

define void @queryPlanDestroy(%TopLevel.ContextType* %contextPtr) {
  ...
}

define private void @staticInit() {
  ...
}

define private void @staticDestroy() {
  ...
}

define internal %TopLevel.ResultRow* 
	@TopLevel.FetchRow(%TopLevel.ContextType* %contextPtr) {
  ...
}
\end{lstlisting}

This is the general structure of the module with the global variables and functions common to all the query plans. Let's see them one by one:
\begin{enumerate}
\item \textsf{\small \%TopLevel.ContextType}: First let's talk about contexts. A context is a set of variables that will be visible to a particular part of the execution of a query plan. Thinking of it as an object-oriented programming model, a context would be the members of a class.
In the execution of a query plan there are two contexts:
\begin{itemize}
\item The global context, which contains the general variables of the query plan. There is a single global instance of this context.
\item The context of the thread, which contains the local variables of each execution thread. There are as many instances of this context as there are threads in the execution of a query plan.
\end{itemize}

The variable\cod{\% TopLevel.ContextType} then represents the data type of the global context variables.

\item \textsf{\small\%TopLevel.ResultRow}: The top-level in this case refers to the root of the execution tree. The rows returned by this node are going to be the rows of the result of the query. This structure represents the format of that row.
For example, in a query like\cod{SELECT x, COUNT (x) FROM table}, if we assume that x is of type int8, the\cod{TopLevel.ResultRow} will have the type <int8, int64> representing x and COUNT respectively .

\item  \textsf{\small @contextInstance}: It is the only instance of the global context, as explained above. This instance will be initialized when the query plan is initialized.

\item \textsf{\small @output.[n]}: These global variables are very important: they are the way to pass the rows ``outside'' of the LLVM module. Each of these variables will have the value of a column of the row that is being outputted. The reason for having a variable for each column is that LLVM has a function that, given a global variable in a module, returns a pointer to it in the module's execution space. In this way, we will be able to inspect the values from the outside.

\item \textsf{\small @initializeContext\_TopLevel.Context}: This is the ``constructor'' of the global context. It creates the instance of the global context and initializes the variables.

\item \textsf{\small @queryPlanInit}: It is the constructor of the query plan. Among other things, it will call\cod{@initializeContext_TopLevel.Context}. This function must be called from the outside to initialize the module.

\item \textsf{\small @queryPlanDestroy}: Destructor of the query plan. Called when you reach EOF and the query plan is not needed anymore.

\item \textsf{\small @staticInit} and\cod{@staticDestroy}: These are functions called by the\cod{@queryPlanInit} to set static variables (which are global outside of the context).

\item \textsf{\small @TopLevel.FetchRow}: This is the MOST important function. Each time it is called, it calculates the next row of the result of the query. Returns NULL when it reaches EOF.
\end{enumerate}

All the modules will have this basic structure.

\subsection{Compilation and Execution}

Once the LLVM module is generated, the ExecutionEngine is created and the query is compiled. All this happens in a line of code:
\bcod
std::unique_ptr <ResultCursor> cursor = 
	queryPlanIR.compile (engine, compileIRBenchmark);
\end{lstlisting}
The cursor will be the object that allows us to traverse the rows one by one.
The module interacts with ``the outside'' (that is, everything that is not compiled by LLVM) through its public functions, which in this case are three:
\begin{enumerate}
\item The initialization function.
\item The FetchRow function.
\item The termination function.
\end{enumerate}

In addition, there is a global variable for each output column, and these variables can also be accessed from the outside. In pseudocode, this is how a query is executed:
\bcod
cursor = queryPlanIR.compile(engine, benchmark)
// The init function is called automatically
while cursor.next():
  row = cursor.values()
  // Process row[0], row[1], etc
cursor.delete()
\end{lstlisting}

\subsection{Translation of Operators to IR}

To translate an operator from AR to IR, each operator has a Builder associated with it. For example, the LIMIT operator has a LimitBuilder associated with it, which is responsible for the translation.
Let's see the example of this particular operator. I use this operator because it is the simplest.
We can think of the translation of an operator as a piece of code that does something with its input and generates an output. Let's take as example the query\sql{SELECT a, b FROM table LIMIT 10}.

In the case of the LIMIT operator, it receives as input the values of columns a and b, and has the same values as output (that is, it does not make any transformation). The two SCAN operators, FULL SCAN and INDEX SCAN, are special since they do not have any input.
If we think of it as a pseudocode, we can implement the two operators independently as follows:
\bcod
// SCAN
class SCAN(tabla):
  def nextRow():
    tabla.next()
    if tabla.inEOF():
      yield EOF
    else:
      yield tabla['a'], tabla['b']
// LIMIT
class LIMIT(childOperator):
  count = 0

  def nextRow():
    while count < 10:
      (a, b) = childOperator.nextRow()
      count += 1
      yield a, b
    yield EOF
\end{lstlisting}

The interesting part of all this is that calling the function\cod{nextRow()}, make yield, etc., has 2 major problems:
\begin{enumerate}
\item It is not very friendly with the instruction cache of the CPU, since it has to make calls to virtual functions.
\item It needs memory to store the intermediate values.
\end{enumerate}

This makes the execution of the query suboptimal, since there is a lot of overhead. We will have to face this problem. Like the fantastic twins joining their rings, these two operators can also be joined in a super operator that we will call Query Plan:
\bcod
class QUERY_PLAN(tabla):
  count = 0

  def nextRow():
    while count < 10:
      tabla.next()    
      if tabla.inEOF():
        yield EOF
      else:
        a, b = tabla['a'], tabla['b']
      
      count += 1
      yield a, b

    yield EOF
\end{lstlisting}

Note that the values of the columns flow ``from bottom to top'' (seeing the query plan as a tree), that is, the values are generated in the leaves and go to the root. This is the main idea of the paper ``Efficiently compiling efficient query plans''~\cite{neumann2011efficiently}. The traditional approach is the opposite, with the problems that we described earlier.

Now let's look directly at the code that translates the LIMIT operator to IR, which is the simplest of all. I put the code here and then I will be commenting part by part. The following code is in the function\cod{LimitOperatorBuilder::_ generateOperatorFunction}, and I put it with color so it looks nicer:
\bsmallcod
void LimitOperatorBuilder::_generateOperatorFunction(
	QueryPlanBuilder& builder, 
	QueryContext& queryContext, 
	llvm::Value* contextInstance, 
	bool /*inlined*/)
{
    const LimitOperator* limitOperator = 
    	dynamic_cast<const LimitOperator*>(this->_operator);
    _zero = builder.getConstant<LimitOperator::limit_type>(0);
    _counterVariable = queryContext.registerContextVariable(
        this->prefix() + ".currentPosition",
        builder.getType<LimitOperator::limit_type>(),
        _zero);

    llvm::Constant* limit = builder.getConstant<LimitOperator::limit_type>(limitOperator->limit);
    llvm::BasicBlock* produceChildBlock = builder.CreateBasicBlock("produceValues");
    llvm::BasicBlock* returnRowBlock = builder.CreateBasicBlock("returnRow");
    llvm::BasicBlock* reachedEOFBlock = builder.CreateBasicBlock("reachedEOF");
    // Operator begin
    // If we have outputted enough rows, finish.
    llvm::Value* counterVariablePtr = queryContext.getVariablePtr(contextInstance,
    		_counterVariable, builder);
    llvm::Value* currentCounter = builder.CreateLoad(counterVariablePtr, "currentCounter");
    llvm::Value* reachedLimit = builder.CreateGenericCmp(QueryPlanBuilder::CMP_GE, 
    		currentCounter, limit, "reachedLimit");
    builder.CreateCondBr(reachedLimit, reachedEOFBlock, produceChildBlock);

    builder.SetInsertPoint(produceChildBlock);
    {
        // Create the function to produce the child values
        OperatorBuilder::OperatorFunction childProduceFunction = this->childBuilder->
        		buildInlinedOperatorFunction(builder, queryContext, contextInstance);
        OperatorBuilder::OperatorFunction::OperatorFunctionResult childResult = 
        		childProduceFunction.invokeOperator(builder);
        // Increment the counter
        llvm::Value* next = builder.CreateAdd(currentCounter, 
        		builder.getConstant<LimitOperator::limit_type>(1));
        builder.CreateStore(next, counterVariablePtr);
        // Generate the requested expressions. 
        // This operator is not able to generate expressions on its own, so it
        // just copies the expressions from the child node.
        for (const SQLExpression& expression : this->_operator->expressions)
            this->setOutputValue(builder, expression, childResult.expressionResolver.
            		getResolvedExpression(expression, builder));
        builder.CreateCondBr(childResult.inEOF, reachedEOFBlock, returnRowBlock);
    }
    builder.SetInsertPoint(returnRowBlock);
    {
        this->returnRow(builder);
    }
    builder.SetInsertPoint(reachedEOFBlock);
    {
        this->returnEOF(builder);
    }
}
\end{lstlisting}

As you will see it is not terribly complicated. Although this is the easiest operator, by far!
Let's see what is happening, and take the opportunity to introduce the concepts.
\bsmallcod
    const LimitOperator* limitOperator = 
    	dynamic_cast<const LimitOperator*>(this->_operator);
    // Nothing crazy here. You only get a pointer to the operator itself.
    _zero = builder.getConstant<LimitOperator::limit_type>(0);
    _counterVariable = queryContext.registerContextVariable(
        this->prefix() + ".currentPosition",
        builder.getType<LimitOperator::limit_type>(),
        _zero);
\end{lstlisting}

Here begins the interesting part. There is a reference to context variables. To understand it, you have to keep in mind that the same portion of the code can be executed in several processors in parallel. The context variables are then the variables of each execution thread. Each operator will have a pointer available to its context, which in turn will have a set of variables specific for each operator.
In this case, the LIMIT operator is saying ``I need the context to store a variable of type\cod{LimitOperator::limit_type} (alias for\cod{uint64_t}), and call it\cod{LimitOperator.CurrentPosition}. The variable must be initialized with the value 0''.
According to the pseudocode that we showed before, this is neither more nor less than the line that says ``count = 0'', with the addition that this is an instance variable of the class, and not a local variable.
\bsmallcod
    llvm::Constant* limit = builder.getConstant<LimitOperator::limit_type>(limitOperator->limit);
    // Here the value for which you are limiting yourself is loaded in a constant (in this case, 10).
    llvm::BasicBlock* produceChildBlock = builder.CreateBasicBlock("produceValues");
    llvm::BasicBlock* returnRowBlock = builder.CreateBasicBlock("returnRow");
    llvm::BasicBlock* reachedEOFBlock = builder.CreateBasicBlock("reachedEOF");
\end{lstlisting}

Here the BasicBlock that the operator will use are created. If you do not know what a BasicBlock is, go right now and read the LLVM tutorial.
\bsmallcod
    llvm::Value* counterVariablePtr = queryContext.getVariablePtr(contextInstance, 
    		_counterVariable, builder);
    llvm::Value* currentCounter = builder.CreateLoad(counterVariablePtr, "currentCounter");
    llvm::Value* reachedLimit = builder.CreateGenericCmp(QueryPlanBuilder::CMP_GE, 
    		currentCounter, limit, "reachedLimit");
    builder.CreateCondBr(reachedLimit, reachedEOFBlock, produceChildBlock);
\end{lstlisting}

This reads the value of the environment variable counter (remember: in LLVM, a variable and its value are two different things! A variable is simply a pointer to a memory address), and compares it with the value of the limit. If it is greater or equal, it jumps to the block that returns EOF, otherwise it jumps to the block that will produce the values of the row:
\bsmallcod
if counter >= limit:
  goto eof
else:
  goto produceChild

    builder.SetInsertPoint(produceChildBlock);
    {
\end{lstlisting}

%Acá inyectamos el código que va a pedirle al operador hijo (en este caso el FULL SCAN) que produzca filas.
Here we inject the code that will ask the child operator (in this case the FULL SCAN) to produce rows.

\bsmallcod
        // Create the function to produce the child values
        OperatorBuilder::OperatorFunction childProduceFunction = this->childBuilder->
        		buildInlinedOperatorFunction(builder, queryContext, contextInstance);
        OperatorBuilder::OperatorFunction::OperatorFunctionResult childResult = 
        		childProduceFunction.invokeOperator(builder);
\end{lstlisting}

Perhaps the name\cod{buildInlinedOperatorFunction} is contradictory: the code is either inline, or a separate function. Sorry about that. In this case, the code that produces the rows is injected directly into the current function.
The struct\cod{OperatorFunctionResult} allows you to abstract from where the child operator values come from. As I mentioned before, the child's code can be generated inline or in a separate function. This structure makes this transparent.
The two most important parts of this structure are:
\bsmallcod
struct OperatorFunctionResult {
    llvm::Value* inEOF = nullptr;
    ExpressionResolver expressionResolver;
    ...
}
\end{lstlisting}

\textsf{\small inEOF} is simply a Boolean value that indicates whether the child returned EOF or not. \cod{expressionResolver} works like a dictionary and allows to bring the values of the row that was generated.
Continuing with the code:
\bsmallcod
        // Increment the counter
        llvm::Value* next = builder.CreateAdd(currentCounter, 
        		builder.getConstant<LimitOperator::limit_type>(1));
        builder.CreateStore(next, counterVariablePtr);
\end{lstlisting}

%Esto es bastante trivial. Incrementamos el contador actual en 1 y almacenamos el resultado en la variable counter.
This is quite trivial. We increase the current counter by 1 and store the result in the counter variable.

\bsmallcod
        // Generate the requested expressions. 
        // This operator is not able to generate expressions on its own, so it
        // just copies the expressions from the child node.
        for (const SQLExpression& expression : this->_operator->expressions)
            this->setOutputValue(builder, expression, childResult.expressionResolver.
            		getResolvedExpression(expression, builder));
\end{lstlisting}

This piece of code ``records'' the output values of this operator. Here basically it copies the values from input to output. Note that when I say that it copies the values, it does not mean that it makes a true copy of the values. In this particular case, the output values are stored in the same registers as the input values. This means that there will be zero overhead for passing the values of the child operator to the parent operator (if there were any, in this particular case there is not). This is the most important concept of all!

\bsmallcod
        builder.CreateCondBr(childResult.inEOF, reachedEOFBlock, returnRowBlock);
    }
\end{lstlisting}

It checks whether the child operator returned EOF. In that case, this operator will also return EOF, otherwise it will ``issue'' the current row.

Two questions can be asked:
\begin{enumerate}
\item Why do we check for EOF after obtaining the values of the child operator?
\item What happens if the child arrived at EOF? What values will be put in the records that we generate?
\end{enumerate}

The answers are:
\begin{enumerate}
\item The most logical thing would be to first check whether EOF was reached and then get the values of the child operator. However, it is done in the other order for simplicity, since checking whether the child arrived at EOF implies an additional if. It is not an optimization (since the compiler should automatically realize this), but rather to make the code simpler.
\item If the child arrived at EOF, the generated row will have garbage and that garbage will be copied to this operator. Since we are going to return EOF, it really does not matter if the generated row has garbage.
\end{enumerate}

\bsmallcod
    builder.SetInsertPoint(returnRowBlock);
    {
        this->returnRow(builder);
    }
\end{lstlisting}

This is the code that indicates that a row must be generated. It is simply a jump to the ``parent operator''. This is the equivalent of the yield.
\bsmallcod
    builder.SetInsertPoint(reachedEOFBlock);
    {
        this->returnEOF(builder);
    }
}
\end{lstlisting}

As before, the EOF indication is made by simply jumping to a special block that transfers the EOF to the parent.
And this is basically what an operator does: configure its context, receive the rows of the child operator, and generate the output for the parent operator.

% Section 5
%!TEX root = snel.tex

\section{Final Remarks}

\subsection{Discussion}

If I had to do everything from scratch, here is what I would do:
\begin{enumerate}
\item I would not make\cod{sqlite3-grandata} a dependency. In fact, it would not make it so dependent on SQL, since there are some queries that are difficult to express in SQL. For this we would have to make a DSL and a parser, which can be quite cumbersome. Or use Lua as an embedded language to express queries to the base.
\item Definitely not in C ++. There is a lot of boilerplate code that only serves to satisfy the gods of C ++. I would do it in Scala. The problem is that there are no bindings of LLVM for Scala, but it is not that difficult to do (see here).
\item I would eliminate the indexes and make the columns compressed. This would imply that all queries would be FULL SCAN, but they would be sequential rather than random scans and in most cases the queries would be faster.
Perhaps a good idea would be to read directly files with the Parquet format, which already have compression included.
\item I would do it distributed, although this is very hard to do. Maybe with Zookeper or something that allows synchronization. This would allow to alleviate the problem of FULL SCAN.
\item I would do it client-server. Nowadays it is imbibed and this generates many problems and inconveniences.
\end{enumerate}

\subsection{Summary}

In this paper, we presented Snel, a relational database engine featuring Just-In-Time (JIT) compilation of queries and columnar data representation. This engine was designed for fast on-line analytics by leveraging the LLVM compiler infrastructure. 

Unlike traditional database engines, it does not provide a client-server interface. Instead, it exposes its interface as an extension to SQLite, for a simple interactive usage from command line and for embedding in applications.
Since Snel tables are read-only, it does not provide features like transactions or updates. This allows queries to be very fast since they don’t have the overhead of table locking or ensuring consistency.

%-----------------------------------------------------
% References
%-----------------------------------------------------

\bibliographystyle{ieeetr}

\bibliography{bib_snel,GD_works}{}

\end{document}